\documentclass[proceedings]{JHEP3}
\PrHEP{PrHEP hep2001}
\conference{International Europhysics Conference on HEP}
\usepackage{epsfig}

\title{Damping of inhomogeneities in neutralino dark matter}
\author{\speaker{Dominik J. Schwarz} \\  
        Institut f\"ur Theoretische Physik, Technische Universit\"at Wien,\\
        Wiedner Hauptstra\ss e 8--10, A-1040 Wien, Austria \\ 
        E-mail: \email{dschwarz@hep.itp.tuwien.ac.at}}
\author{Stefan Hofmann and Horst St\"ocker \\
        Institut f\"ur Theoretische Physik, Universit\"at Frankfurt, \\
        Postfach 11 19 32, D-60054 Frankfurt am Main, Germany \\ 
        E-mail: \email{stehof@th.physik.uni-frankfurt.de}}

\abstract{
The lightest supersymmetric particle, most likely the neutralino, might 
account for a large fraction of dark matter in the Universe. We show that the 
primordial spectrum of density fluctuations in neutralino cold dark matter 
(CDM) has a sharp cut-off due to two damping mechanisms: collisional damping 
during the kinetic decoupling of the neutralinos at ${\cal O}(10 \mbox{\ MeV})$
and free streaming after last scattering of neutralinos. The cut-off in the 
primordial spectrum defines a minimal mass for CDM objects in hierarchical 
structure formation. For typical neutralino and sfermion masses the first 
gravitationally bound neutralino clouds have masses above $10^{-6} M_\odot$.}

\begin{document}

\section{Introduction}

The evidence that there is a significant amount of cold dark matter (CDM) in 
the Universe has been confirmed by several measurements recently. An analysis 
of the temperature anisotropies of the cosmic microwave background (CMB) gives  
for the mass density of CDM $\Omega_{\rm cdm}h^2 = 0.12\pm 0.05$ (weak priors)
\cite{Netterfield}, while from the same analysis the mass density of baryons 
is much smaller, $\Omega_{\rm b}h^2 = 0.021^{+0.004}_{-0.003}$. The latter is 
consistent with the prediction of 
primordial nucleosynthesis \cite{Tytler}. Large galaxy redshift surveys agree 
with the CMB measurements. From the analysis of over 160,000 galaxies the 
authors of Ref.\ \cite{Percival} find $(\Omega_{\rm cdm}+\Omega_{\rm b})h = 
0.20\pm 0.03$ and $\Omega_{\rm b}/(\Omega_{\rm cdm} + \Omega_{\rm b}) = 
0.15\pm 0.07$. These recent findings agree well with the picture of 
hierarchical structure formation that emerged over the years from the study 
of much smaller structures as galaxies, groups of galaxies and clusters. An 
up to date summary of the evidence for CDM is presented in \cite{Turner}.  

The characteristic feature of CDM is its non-relativistic equation of state, 
i.e. $P \ll \rho$, in order to ensure that large scale structure has enough 
time to form. The behaviour of all CDM candidates is the same at large scales 
($> 1$ Mpc), whereas at the galactic scale and below various CDM candidates
might be distinguishable. To learn more about the nature of CDM we study 
the small-scale structure of CDM (see also \cite{Gurevich}). 

One of the most popular CDM candidates is the neutralino $\tilde{\chi}^0_1$, 
which probably is the lightest supersymmetric particle. In the context of the 
constrained minimal supersymmetric standard model it is almost a pure bino 
with mass $M_{\tilde{\chi}} = {\cal O}(100 \mbox{\ GeV})$ \cite{Ellis}.

In this contribution we discuss the two damping mechanisms, collisional damping
during the kinetic decoupling of neutralinos from the radiation fluid and 
free streaming thereafter, that are relevant for the formation of the very 
first structures of neutralino CDM. For more details see \cite{HSS}.  

\section{Chemical and kinetic decoupling}

Dark matter particles that are massive and subject to weak interactions 
decouple chemically long before they decouple kinetically from the radiation
fluid (photons and leptons) \cite{SSW,HSS}. This can be understood by comparing
the neutralino annihilation rate $\Gamma_{\rm ann}(\tilde\chi + \tilde\chi 
\rightarrow l + \bar l) \equiv \langle v\sigma_{\rm ann}\rangle n_{\tilde\chi}$
to the rate of elastic scatterings $\Gamma_{\rm el}(\tilde\chi + l \rightarrow 
\tilde\chi + l) \equiv \langle v\sigma_{\rm el}\rangle n_{\rm l}$. Since 
$M_{\tilde\chi} \gg T$ at chemical decoupling (freeze out), the number density 
of neutralinos $n_{\tilde\chi}$ is suppressed with respect to the number 
density of the relativistic leptons $n_{\rm l}$. Therefore the chemical 
decoupling temperature $T_{\rm cd}$ is much larger than the temperature of 
kinetic decoupling $T_{\rm kd}$. The dominant contribution to the elastic
scattering amplitudes for bino-like neutralinos comes from slepton exchange.
It was shown in \cite{CKZ} that the contribution of neutralino-photon 
scattering is negligible. To estimate the temperature of kinetic 
decoupling we have to calculate the relaxation time $\tau$. The momentum 
transfer per elastic scattering is $\Delta p_{\tilde\chi}/p_{\tilde\chi} 
\sim T/M_{\tilde\chi} \ll 1$, thus we need many scatterings to change the 
momentum of an individual CDM particle by a significant amount. Kinetic 
decoupling occurs when the relaxation time starts to exceed the Hubble time,
\begin{equation}
\tau 
= \frac{p_{\tilde\chi}}{\Delta p_{\tilde\chi}} \frac 1{\Gamma_{\rm el}} 
\sim \frac 1 H , 
\end{equation}
from which we can determine $T_{\rm kd}$. Figures \ref{fig1} and \ref{fig2} 
show the dependence of the decoupling temperatures on the sfermion mass. 
\DOUBLEFIGURE{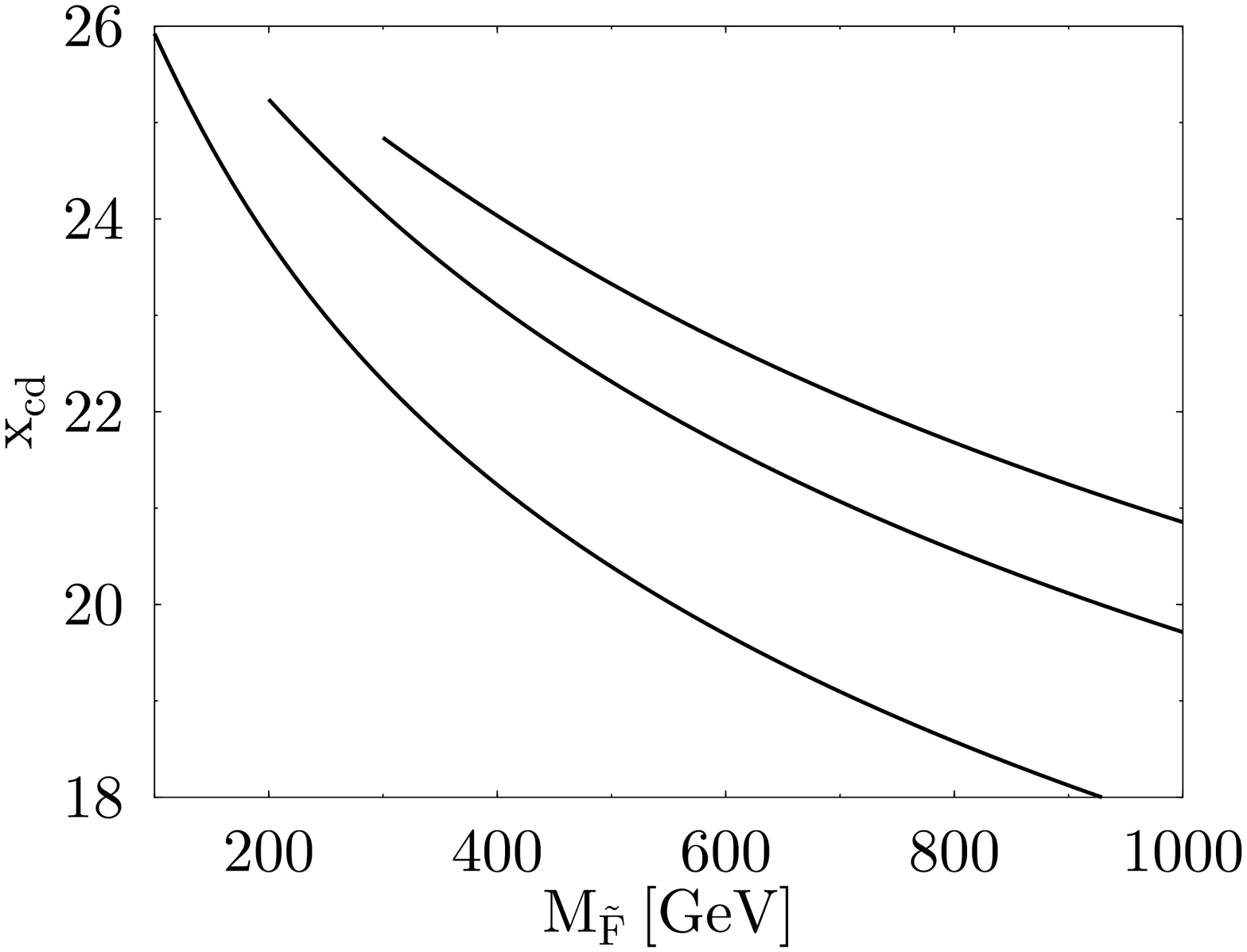,width=\linewidth}{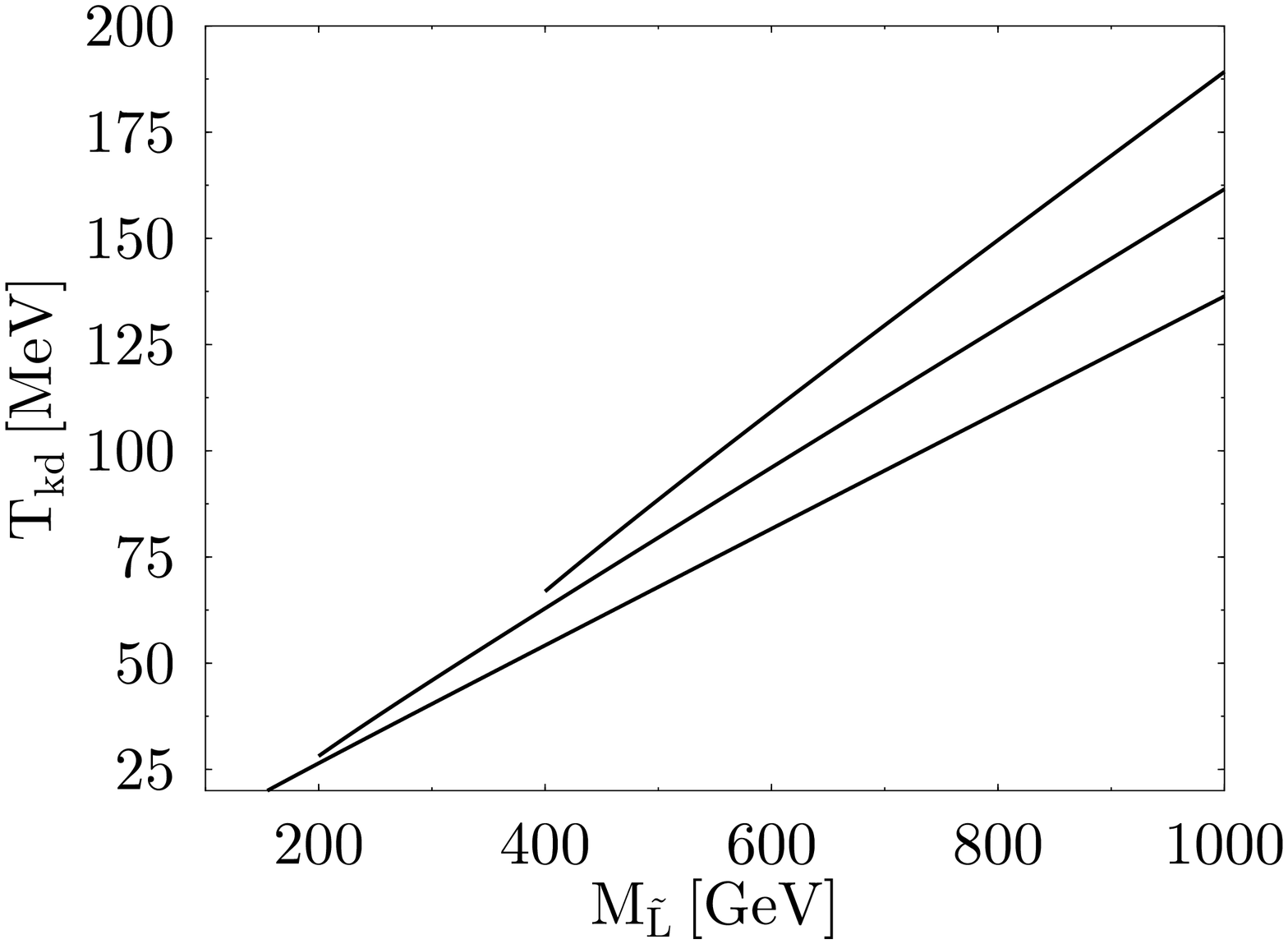,width=\linewidth}
{Chemical decoupling. $x_{\rm cd} \equiv M_{\tilde\chi}/T_{\rm cd}$ as a 
function of the sfermion mass for $M_{\tilde\chi} = 50, 100, 200$ GeV 
(bottom to top).\label{fig1}}
{Kinetic decoupling. $T_{\rm kd}$ as a function of the slepton mass for 
$M_{\tilde\chi} = 50, 100, 200$ GeV (bottom to top).
\label{fig2}}
For reasonable masses $M_{\tilde\chi} = 150$ GeV and $M_{\tilde F} = 
M_{\tilde L} = 250$ GeV we find $T_{\rm cd} \approx 6$ GeV and 
$T_{\rm kd} \approx 40$ MeV.  

\section{Collisional damping}

During the process of kinetic decoupling the neutralinos acquire a finite 
mean free path. Density inhomogeneities on scales of the diffusion 
length are damped by the mechanism of collisional damping. It is convenient to
describe the CDM as an imperfect fluid. We have shown in \cite{HSS} that 
the coefficient of heat conduction vanishes at the leading order in 
$T/M_{\tilde\chi}$. Thus the dominant contribution to collisional damping comes
from bulk and shear viscosity. Since energy of the CDM fluid can be transferred 
to the radiation fluid, which acts here like an inner degree of freedom for the 
CDM particles, the bulk viscosity does not vanish. Nevertheless, the radiation
fluid can be treated as a perfect fluid since $\rho_{\rm rad} \gg 
\rho_{\rm cdm}$ at kinetic decoupling of the neutralinos. We calculated the 
relevant coefficients of transport from kinetic theory in \cite{HSS}. At linear
order in the relaxation time the coefficients of shear and bulk viscosity 
become $\eta \approx n_{\tilde\chi} T \tau$ and $\zeta \approx 5\eta/3$, 
respectively. 

The density inhomogeneities in CDM are damped exponentially below the scale 
$M_{\rm d}$ due to viscosity \cite{Weinberg,HSS} 
\begin{equation}
\left({\delta\rho_{\tilde\chi} \over \rho_{\tilde\chi}}\right)_k \propto 
\exp\left[-\frac 3 2 \int_0^{t_{\rm kd}} 
\frac{T \tau}{M_{\tilde\chi}}\; k_{\rm ph}^2 {\rm d}t
\right] = \exp\left[ - \left(\frac{M_{\rm d}}{M}\right)^{2/3}\right],
\label{cd}
\end{equation}
where it is useful to work with the CDM mass enclosed in a sphere of radius
$2\pi/k_{\rm ph}$, because the mass in CDM is time independent. In figure 
\ref{fig3} we plot the damping mass $M_{\rm d}$ as a function of the neutralino
mass for various values of the slepton mass. The damping (\ref{cd}) 
provides a small-scale cut-off in the primordial spectrum of density 
perturbations in neutralino CDM.  
\EPSFIGURE{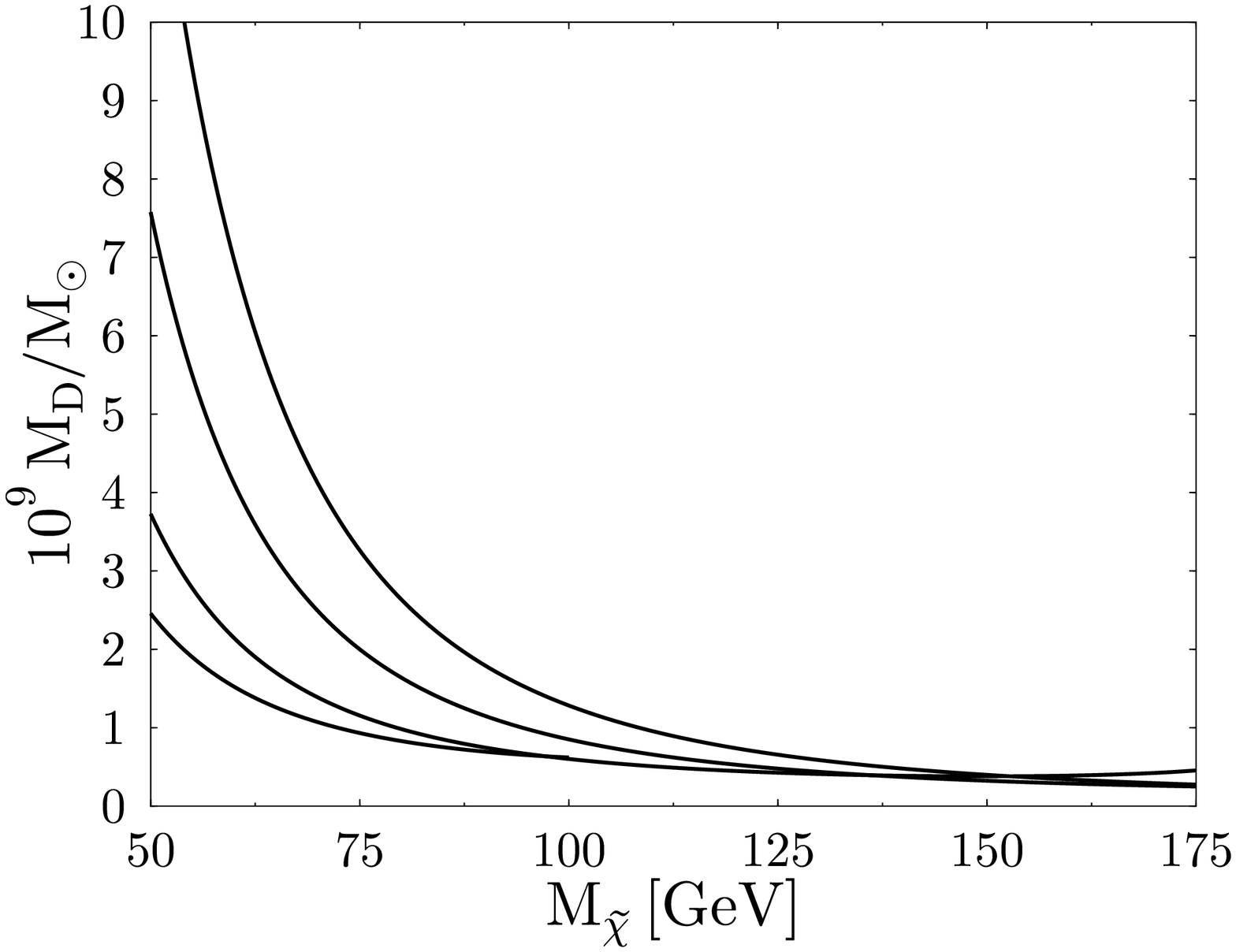,width=0.6\linewidth}
{Mass scale of collisional damping of CDM inhomogeneities
against the neutralino mass for $M_{\tilde l} = 150,200,300,400$ GeV
(bottom to top).\label{fig3}}

\section{Free streaming}

Once the temperature in the Universe drops below $T_{\rm kd}$ the rate of 
elastic scatterings is not high enough to keep the neutralinos in thermal 
equilibrium with the radiation fluid. The neutralinos enter the regime of
free streaming. This process continues to smear out inhomogeneities, since 
the individual neutralinos do not move coherently. In a forthcoming paper 
we will show by means of kinetic theory that the damping due to free 
streaming goes as
\begin{equation}
\left({\delta\rho_{\tilde\chi} \over \rho_{\tilde\chi}}\right)_k \propto
\exp\left[- \frac{T_{\rm kd}}{2 m_{\tilde\chi}} \left(\frac{k_{\rm ph}}{H}
\right)^2_{T=T_{\rm kd}} \ln^2 \left(\frac{a}{a_{\rm kd}}\right)\right] = 
\exp\left[ - \left(\frac{M_{\rm fs}(a)}{M}\right)^{2/3}\right].
\label{fs}
\end{equation} 
The mass scale of damping from free streaming $M_{\rm fs}$ is written as
a function of the cosmic scale factor $a$. The damping scale grows 
logarithmically with the scale factor. This calculation agrees with the 
estimate of the free streaming scale from the free streaming length as 
presented in \cite{HSS} up to a numerical factor $(2\pi/\sqrt{6})^3 \approx 
17$. (We previously underestimated the free streaming mass by that factor.) 
The ratio
\begin{equation}
\label{mfs}
\frac{M_{\rm fs}}{M_{\rm d}} = \left[\sqrt{\frac 5 3}\ln \frac{a}{a_{\rm kd}}
\right]^3
\end{equation}
exceeds unity for $a > 2.2 a_{\rm kd}$, thus free streaming starts to dominate 
the damping from collisional damping once the Universe has doubled its size 
after kinetic decoupling. It is interesting to evaluate (\ref{mfs}) at the
time of matter-radiation equality, since this is the moment when CDM density
perturbations start to grow linearly with the expansion. For a kinetic 
decoupling temperature $T_{\rm kd} = 40$ MeV and for $(\Omega_{\rm cdm} + 
\Omega_{\rm b}) h^2 = 0.15$ we find $M_{\rm fs}(a_{\rm eq})/M_{\rm d} 
\approx 1.3 \times 10^4$, thus $M_{\rm fs}(a_{\rm eq}) \approx 5 \times 10^{-6} 
M_\odot$ for $M_{\tilde\chi} = 150$ GeV. 

Typically the free streaming mass at the time of equality is of the order
of $10^{-6} - 10^{-5} M_\odot$, which is in striking contrast to claims
in the literature (see e.g.~\cite{Gurevich}) that the minimal mass for the 
first objects would be $\sim 10^{-13} (150 \mbox{\ GeV}/M_{\tilde\chi})^3 
M_\odot$. The huge difference to our result comes mainly from the false
assumption that kinetic decoupling happens simultaneously with chemical 
decoupling. 

\section{Conclusions}

We have shown that collisional damping and free streaming smear out all 
power of primordial density inhomogeneities in neutralino CDM below
$\sim 10^{-6} M_\odot$. This implies that there is a peak (subhorizon CDM 
density perturbations grow logarithmically during the radiation epoch) in the
power spectrum close to the cut-off and therefore we have found the minimal
mass for the very first objects, if CDM is made of neutralinos. Our result 
does not depend in a strong way on the parameters of the supersymmetric  
standard model. 

According to the picture of the hierarchical formation of structure these very
small first objects are supposed to merge and to form larger objects,
eventually galaxies and larger structures. It is unclear whether some of 
the very first objects have a chance to survive. CDM simulations show structure 
on all scales, down to the resolution of the simulation \cite{K}. 
However, the dynamic range is not sufficient to deal with the first CDM 
objects, so the fate of the first CDM clouds is an open issue. A cloudy 
distribution of CDM in the galaxy would have important implications for 
direct and indirect searches for dark matter.

\acknowledgments

D.J.S. thanks Francesco Vissani for references to the literature and 
acknowledges financial support of the Austrian Academy of Sciences.

\end{document}